\documentclass[12pt]{article}
\usepackage{a4,amsfonts,amssymb,amsmath,amsthm,graphicx}

\newcommand{\BB}{\vert_{\partial M}}
\newcommand{\tr}{{\rm tr}\,}
\newcommand{\Dir}{\widehat D}
\newcommand{\iM}{\int_M d^nx\sqrt{g}}
\newcommand{\idM}{\int_{\partial M} d^{n-1}x\sqrt{h}}

\begin{document}

\title{Chiral anomaly for local boundary conditions }

\author{Valery N.Marachevsky \thanks{email: maraval@mail.ru,
root@VM1485.spb.edu} \\
{\it V. A. Fock Institute of Physics, St. Petersburg
University,}\\
{\it 198504 St. Petersburg, Russia}\\[10pt]
 Dmitri V.Vassilevich\thanks{Also at V.~A.~Fock Insitute of
Physics, St.Petersburg University, Russia; email:
Dmitri.Vassilevich@itp.uni-leipzig.de}\\
{\it Institut f\"{u}r Theoretische Physik, Universit\"{a}t
Leipzig,}\\
{\it D-04109 Leipzig, Germany } }

\maketitle

\begin{abstract}

It is known that in the zeta function regularization and in the
Fujikawa method chiral anomaly is defined through a coefficient
in the heat kernel expansion for the Dirac operator.
In this paper we apply the heat kernel methods to calculate 
boundary contributions to the chiral anomaly for local (bag)
boundary conditions. As a by-product some new results on the
heat trace asymptotics are also obtained.

\end{abstract}


\section{Introduction}

Chiral anomaly, which was discovered more than 30 years ago
\cite{abj}, still plays an important role in particle physics. On
smooth manifolds without boundaries many successful approaches to
the anomalies exist\footnote{The reader may consult, for example,
ref.\ \cite{books} or the original papers
\cite{Wess:yu,Fujikawa:1979ay,Faddeev:pc}.}. Modern developments
in theoretical physics require anomaly calculations on branes and
domain walls (see, e.g., \cite{Chen:2002bw}). Taking into account
also more traditional applications as the bag model of hadrons
\cite{bag}, we see that understanding the chiral anomaly in the
presence of boundaries or singularities is an important task.

In the case of non-trivial background fields, and especially in
the presence of boundaries or singularities, the heat kernel
technique seems to be the most adequate one for analysing the
one-loop effects (see \cite{Vassilevich:2003xt} for a recent
review). The heat kernel approach to the anomalies is essentially
equivalent to the Fujikawa approach \cite{Fujikawa:1979ay} and to
the calculations based on the finite-mode regularization
\cite{Andrianov:1983fg}, but it can be more easily extended to
complicated geometries.

In the present paper we consider an euclidean version of the bag
boundary conditions \cite{bag} (see also \cite{Symanzik:1981wd}).
Although these are the most simple and physically most natural
boundary conditions, chiral anomaly for these conditions has not
been calculated so far\footnote{There exist calculations of global
chiral anomaly (i.e. for a position-independent chiral
transformation parameter) for some other types of boundary
operators, and also calculation of the scale anomaly for bag
boundary conditions. Literature on these topics is rather large,
so that for uniformity we quote nobody here. A literature survey
can be found in \cite{Vassilevich:2003xt}.}. This can be explained
by the fact that the calculation is indeed rather involved. We use
the zeta function regularization and our approach to the anomaly
follows closely the paper \cite{Blau:1988jp}. Mathematical foundations
for analysing spectral geometry of the Dirac operator with local
boundary conditions were developed in \cite{BG-Dirac}. A simple
overview of spectral properties of Dirac operator on manifolds
with boundary can be found in \cite{GEbook}.

To calculate the anomaly we employ the following strategy.
First we relate the anomaly to a heat kernel coefficient
for the square of the Dirac operator. Then we generalise the
problem and consider the heat kernel for
an arbitrary operator of the Laplace type with mixed boundary
conditions and with a matrix valued smearing function. This generalisation
allows us to use the heavy machinery of the heat kernel expansion
for such operators. In particular, important information follows
from two simple examples of boundary value problems in one and two
dimensions. In this way we are able to calculate first five heat 
kernel coefficients. Chiral anomaly is then obtained simply
by substituting explicit expressions for the connection, the
potential, and other quantities in terms of the background vector and
axial vector fields. 

In the next section we remind some basic formulae regarding the zeta
function regularization, chiral anomaly, and local (bag) boundary 
conditions. Sec.\ 3 is devoted to the relationship between Dirac
and Laplace operators. The heat kernel coefficients are calculated
in sec.\ 4, which is the main technical part of this paper. In 
sec.\ 5 we return to the Dirac operator and finally calculate boundary
contributions to chiral anomaly in two and four dimensions. Our
results and their possible extensions are discussed in sec.\ 6.
Appendix \ref{hkapp} contains the heat kernel coefficients for mixed
boundary conditions with a scalar smearing function. Appendix \ref{partapp}
gives details of the particular case calculations used in sec.\ 4. 

\section{Boundary conditions and chiral anomaly}

Let us consider an $n$-dimensional Riemannian manifold $M$ with
boundary $\partial M$. Dirac $\gamma$-matrices satisfy the
Clifford commutation relation

\begin{equation}
  \{ \gamma^\mu, \gamma^\nu \} =- 2g^{\mu\nu}.
  \label{A1}
\end{equation}

The $\gamma$-matrices defined in this way are anti-hermitian,
$\gamma^{\mu \dag} = -\gamma^\mu$. We also need the chirality
matrix which will be denoted $\gamma^5$ independently of the
dimension of $M$. As usual,  $\gamma^5\gamma^\mu =-\gamma^\mu
\gamma^5$ and $\gamma^{5\dag}=\gamma^5$. We assume that $n$ is
even. We fix the sign of $\gamma^5$ by choosing
\begin{equation}
\gamma^5=\frac{i^{n(n+1)/2}}{n!} \epsilon^{\mu\nu\dots\rho}
\gamma_\mu\gamma_\nu\dots\gamma_\rho \,.\label{defg5}
\end{equation}

Consider the Dirac operator
\begin{equation}
  \Dir =\gamma^\mu\left(\partial_\mu+ V_\mu+ iA_\mu\gamma_5- \frac{1}{8} [ \gamma_\rho,
  \gamma_\sigma ] \sigma^{[\rho \sigma]}_\mu \right)
  \label{A2}
\end{equation}
in external vector $V_\mu$ and axial vector $A_\mu$ fields. We
suppose that $V_\mu$ and $A_\mu$ are anti-hermitian matrices in
the space of some representation of the gauge group.
$\sigma^{[\rho \sigma]}_\mu$ is the spin-connection\footnote{The
spin-connection must be included even on a flat manifold if the
coordinates are not Cartesian.}.

The Dirac operator transforms covariantly under infinitesimal
local gauge
transformations:
\begin{eqnarray}
&&\delta_\lambda A_\mu =[A_\mu ,\lambda ] \nonumber \\
&&\delta_\lambda V_\mu = \partial_\mu \lambda +[V_\mu ,\lambda ]
\nonumber \\
&&\Dir \to \Dir +[\Dir ,\lambda ] \label{gauge}
\end{eqnarray}
and under infinitesimal local chiral transformations:
\begin{eqnarray}
&&\tilde\delta_\varphi A_\mu =\partial_\mu \varphi +[V_\mu
,\varphi ],
\nonumber \\
&&\tilde\delta_\varphi V_\mu =-[A_\mu ,\varphi ] ,\nonumber \\
&&\Dir \to \Dir + i\{ \Dir , \gamma^5 \varphi \} \,.
\label{chiral}
\end{eqnarray}
The parameters $\lambda$ and $\varphi$ are anti-hermitian
matrices.

We adopt the zeta-function regularization\footnote{Our approach to
the effective action for fermions is close to that in the papers
\cite{Blau:1988jp,Wipf:1994dy}. We refer to these works for a more
detailed derivation of chiral anomaly (with somewhat more accurate
treatment of the zero mode problem).} and write the effective
action for the Dirac fermions as
\begin{equation}
W=-\ln \det \Dir = -\frac 12 \ln \det \Dir^2 = \frac 12 \zeta'(0)
+\frac 12 \ln (\mu^2) \zeta (0) \,,\label{detD}
\end{equation}
where
\begin{equation}
\zeta (s)={\rm Tr} ( \Dir^{-2s} )\,, \label{zeta}
\end{equation}
prime denotes differentiation with respect to $s$, and ${\rm Tr}$
is the functional trace.

It is easy to show that the effective action (\ref{detD}) is gauge
invariant, $\delta_\lambda W=0$, and that the variation of $W$
under an infinitesimal chiral transformation reads

\begin{equation}
\mathcal{A}:=\tilde\delta_\varphi W =-2i {\rm Tr} (\gamma^5 \varphi
\Dir^{-2s} )\vert_{s=0} \,. \label{chian}
\end{equation}

Let us define an integrated heat kernel for a second order
elliptic partial differential operator $L$ by the equation:
\begin{equation}
K(Q,L,t):={\rm Tr} \left( Q \exp (-tL) \right) \,,\label{defhk}
\end{equation}
where $Q(x)$ is a matrix valued function. For the boundary conditions
we consider in this paper (see eq.\ (\ref{mixedbc}) below)
there
exists an asymptotic expansion \cite{Gilkey:1994} as $t\to 0$:
\begin{equation}
K(Q,L,t) \simeq \sum_{k=0}^\infty a_k (Q,L) t^{(k-n)/2} \,.
\label{hkexp}
\end{equation}

The heat kernel is related to the zeta function by
the Mellin transformation:
\begin{equation}
{\rm Tr} (\gamma^5 \varphi \Dir^{-2s} )=\Gamma (s)^{-1}
\int_0^\infty dt\, t^{s-1} K(\gamma^5\varphi, \Dir^2, t)
\,.\label{zetahk}
\end{equation}
In particular\footnote{A rather formal way to derive this equation
consists in integration of the asymptotic expansion
(\ref{hkexp}).},
\begin{equation}
\mathcal{A} =-2i a_n (\gamma^5\varphi,\Dir^2) \,.\label{anomhk}
\end{equation}
The same expression for the anomaly follows also from the
Fujikawa approach \cite{Fujikawa:1979ay}.

We shall need some basic notions from differential geometry. Let
$R_{\mu\nu\rho\sigma}$ be the Riemann tensor, and let
$R_{\mu\nu}={R^\sigma}_{\mu\nu\sigma}$ be the Ricci tensor. With
our sign convention the scalar curvature $R=R_\mu^\mu$ is $+2$ on
the unit sphere $S^2$. Curvature does not play any important role
in our calculations. However, we shall see below that curved space
offers no complications in our approach compared to the flat case.

If the manifold $M$ has a boundary, boundary conditions should be
imposed on the spinor field $\psi$. We need several basic
definitions regarding differential geometry of manifolds with
boundary. Let $\{ e_j \}$, $j=1,\dots,n$ be a local orthonormal
frame for the tangent space to the manifold and let on the
boundary $e_n$ be an inward pointing normal vector. Then $\{ e_a
\}$, $a=1,\dots,n-1$ can be identified with a local orthonormal
frame for the tangent space to the boundary. 
The frame $\{ e_j \}$ will be used to transform curved (world)
indices $\mu, \nu ,\dots ,\sigma$
to ``flat'' indices and back. For example, for a vector
$v_\mu$ this transformation reads: $v_j=e_j^\mu v_\mu$,
$v_a=e_a^\mu v_\mu$, $v_n=e_n^\mu v_\mu$. In Euclidean space there is
no distinction between flat upper and lower indices.

The extrinsic
curvature is defined by the equation
\begin{equation}
L_{ab}=\Gamma_{ab}^n \,,\label{Lab}
\end{equation}
where $\Gamma$ is the Christoffel symbol. For example, on the unit
sphere $S^{n-1}$ which bounds the unit ball in $\mathbb{R}^n$ the
extrinsic curvature is $L_{ab}=\delta_{ab}$.

We impose local\footnote{Locality means that the
projector $\Pi_-$ acts at each point of the boundary 
independently. An example of non-local boundary
conditions can be found in \cite{APS}.} boundary conditions:
\begin{equation}
\Pi_-\psi \BB =0,\qquad \Pi_-=\frac 12 \left( 1-\gamma_5 \gamma_n
\right) \,,\label{bagbc}
\end{equation}
which are nothing else than a Euclidean version of the MIT bag
boundary conditions \cite{bag}. For these boundary conditions
$\Pi_-^\dag =\Pi_-$, and the normal component of the fermion
current $\psi^\dag \gamma_n \psi$ vanishes on the boundary.

An important comment on chiral transformations of the boundary
conditions (\ref{bagbc}) is in order. Finite version of the infinitesimal
transformation (\ref{chiral}) reads:
\begin{equation}
\Dir \to \Dir_\varphi = e^{i\varphi\gamma_5} \Dir e^{i\varphi\gamma_5}\,.
\label{finchi}
\end{equation}
This relation yields the following transformation law for the boundary
projector:
\begin{equation}
\Pi_-\to \Pi_-^{[\varphi ]} =e^{-i\varphi\gamma_5} \Pi_- e^{i\varphi\gamma_5}
=\frac 12 \left( 1-\gamma_5\gamma_n e^{2i\varphi \gamma_5} \right) \,,
\label{chipi}
\end{equation}
so that the boundary condition
\begin{equation}
\Pi_-^{[\varphi]} \psi \BB =0 \label{chibag}
\end{equation}
remains consistent with (\ref{bagbc}) and (\ref{finchi}). Eq.\ (\ref{chibag})
represents an Euclidean version \cite{Hrasko:1983sj} of chiral bag boundary
conditions \cite{Chodos:1975ix}. The boundary conditions (\ref{chibag})
are considerably more complicated than (\ref{bagbc}). Even such fundamental
property of (\ref{chibag}) as the strong ellipticity (which ensures, for
example, existence of only simple poles of the zeta function) has been
established only recently \cite{Beneventano:2003hv}. Fortunately, as we 
stay at the level of linear perturbations, the condition (\ref{bagbc}) is
enough for our purposes. However, already the Wess-Zumino consistency
conditions \cite{Wess:yu}
which imply two consequent chiral transformations require
a more general setting of the chiral bag (\ref{chibag}).
\section{Dirac and Laplace operators}

To calculate the anomaly (\ref{anomhk}) it is convenient to
consider a more general problem of calculation of the coefficients
$a_k(Q,L)$ (see eq.\ (\ref{hkexp})) for general matrix valued
function $Q$ and general operator $L$ of Laplace type. Any
operator of Laplace type can be expanded locally as
\begin{equation}
  L = - (g^{\mu\nu} \partial_\mu \partial_\nu + a^\sigma \partial_\sigma + b ),
  \label{B4}
\end{equation}
where $a$ and $b$ are some matrix valued functions. One can always
introduce a connection $\omega_\mu$ and another matrix valued
function $E $ so that $L$ takes the form:
\begin{equation}
  L= - (g^{\mu\nu}\nabla_\mu \nabla_\nu + E)  \label{B5}
\end{equation}
Here $\nabla_\mu $ is a sum of covariant Riemannian derivative
with respect to metric $g_{\mu\nu}$ and connection $\omega_\mu$.
One can, of course, express $E$ and $\omega$ in terms of $a^\mu$,
$b$ and $g_{\mu\nu}$:

\begin{eqnarray}
&&\omega_\mu = \frac{1}{2} g_{\mu\nu} (a^\nu +
  g^{\rho\sigma}\Gamma_{\rho\sigma}^\nu) , \label{B6}
\\
&& E = b - g^{\mu\nu}(\partial_\nu \omega_\mu +
\omega_\mu\omega_\nu
  -\omega_\rho\Gamma_{\mu\nu}^\rho )    \label{B7}
\end{eqnarray}

For the future use we introduce also the field strength for
$\omega$:
\begin{equation}
 \Omega_{\mu\nu}=\partial_\mu\omega_\nu -
\partial_\nu\omega_\mu + [\omega_\mu, \omega_\nu] \,.\label{Omega}
\end{equation}

The connection $\omega_\mu$ will be also used to construct
covariant derivatives. It will be convenient to work with
flat indices (denoted by Latin letters) which we have introduced
in the previous section.
The subscript $;i\dots jk$ will
be used to denote repeated covariant derivatives with the
connection $\omega$ and the Christoffel connection on $M$. The
subscript $:a\dots b c$ will denote repeated covariant derivatives
containing $\omega$ and the Christoffel connection on the
boundary. Difference between these two covariant derivatives is
measured by the extrinsic curvature (\ref{Lab}). For example,
$E_{;ab}=E_{:ab}-L_{ab}E_{;n}$.

Let us now turn to boundary conditions. We assume given two
complementary projectors $\Pi_\pm$, $\Pi_-+\Pi_+=I$ and define
mixed boundary conditions by the relations

\begin{equation}
\Pi_-\psi \BB =0\,,\quad \left( \nabla_n + S\right) \Pi_+ \psi \BB
=0 \,, \label{mixedbc}
\end{equation}
where $S$ is a matrix valued function on the boundary. In other
words, the components $\Pi_-\psi$ satisfy Dirichlet boundary
conditions, and $\Pi_+\psi$ satisfy Robin (modified Neumann) ones.

It is convenient to define
\begin{equation}
\chi =\Pi_+ - \Pi_- \,.\label{defchi}
\end{equation}

Now we have to calculate the geometric quantities introduced above
for generic $L$ in the particular case $L=\Dir^2$. After lengthy
but straightforward calculation one obtains from (\ref{A2}),
(\ref{B6}) -- (\ref{Omega}) (see also \cite{DeBerredo-Peixoto:2001qm}
for the abelian case):
\begin{eqnarray}
&&  \omega_\mu= V_\mu - \frac{i}{2}[\gamma_\mu, \gamma_\nu] A^\nu
\gamma_5 -
  \frac{1}{8}[\gamma_\rho,\gamma_\sigma] \sigma^{[\rho,\sigma]}_\mu  ,
  \label{B9} \\
  &&  E=-\frac{1}{2}\gamma^\mu\gamma^\nu V_{\mu\nu} +
  (n-3) \gamma^\mu\gamma^\nu A_\mu A_\nu - A_\mu A^\mu -
  \frac{1}{4} R + iD_{\mu}A^\mu \gamma_5 ,
  \label{B10}\\
&& \Omega_{\mu\nu}= i \gamma^\kappa
 (D_{\mu}A_{\kappa})\gamma_{\nu}
\gamma_{5} - i \gamma^\kappa
(D_{\nu}A_{\kappa})\gamma_{\mu}\gamma_{5} + i A_{\mu\nu}\gamma_{5}
- [A_{\mu}, A_{\nu}] + V_{\mu\nu}  \nonumber
\\ && \qquad \quad +
\frac{1}{4}\gamma^{\kappa}\gamma^{\tau}R_{\kappa\tau\mu\nu} -
[A_{\mu},A_{\kappa}]\gamma^{\kappa}\gamma_{\nu} +
[A_{\nu},A_{\kappa}]\gamma^{\kappa}\gamma_{\mu} - \nonumber \\
 && \qquad \quad
- \gamma^{\kappa}A_{\kappa}\gamma_{\mu}\gamma^{\tau}
A_{\tau}\gamma_{\nu} +
\gamma^{\kappa}A_{\kappa}\gamma_{\nu}\gamma^{\tau}
A_{\tau}\gamma_{\mu} \label{Om}
\end{eqnarray}
with the notations $V_{\mu\nu} = \partial_{\mu} V_{\nu} -
\partial_{\nu} V_{\mu} + [V_{\mu}, V_{\nu}]$,
$A_{\mu\nu} = D_{\mu} A_{\nu} - D_{\nu} A_{\mu}$, $D_{\mu} A_{\nu} =
\partial_{\mu} A_{\nu} - \Gamma^{\rho}_{\mu\nu}A_{\rho}
+ [V_{\mu},A_{\nu}]$. Note, that the gauge covariant derivative $D_\mu$
differs from $\nabla_\mu$ defined above.

Since $\Dir$ is a first order differential operator it was enough
to fix the boundary conditions (\ref{bagbc}) on a half of the
components. To proceed with a second order operator $L=\Dir^2$ we
need boundary conditions on the remaining components as well. They
are defined by the consistency condition:
\begin{equation}
\Pi_-\Dir \psi \BB =0 \,,\label{conscond}
\end{equation}
which is equivalent to the second (Robin) boundary condition in
(\ref{mixedbc}) with
\begin{equation}
S= - \frac{1}{2}\Pi_+L_{aa} \,.\label{B11}
\end{equation}


\section{Asymptotic expansion of the heat kernel}


\subsection{General strategy}

In this section we study the short $t$ asymptotics (\ref{hkexp})
for an arbitrary operator $L$ of Laplace type. Two particular
cases of the expansion (\ref{hkexp}) are known. The heat kernel
coefficients $a_k$, $k=0,1,2,3,4,5$ for a scalar $Q=fI$ (where $f$
is a function and $I$ is the unit operator) are presented in
Appendix \ref{hkapp}. The case of arbitrary $Q$ but pure Dirichlet
or Neumann boundary conditions (i.e. when either $\Pi_+$ or
$\Pi_-$ is zero) was studied in \cite{Branson:1997ze}. Here we
need a combination of these two cases. Namely, we are interested
in mixed boundary conditions and a matrix valued $Q$.

According to the general theory \cite{Gilkey:1994} the
coefficients $a_k(Q,L)$ are locally computable. This means that
each $a_k(Q,L)$ can be represented as a sum of volume and boundary
integrals of local invariants constructed from $Q$, $\Omega$, $E$,
the curvature tensor, and their derivatives. Boundary invariants
may also include $S$, $L_{ab}$ and $\chi$. Total mass dimension of
such invariants should be $k$ for the volume terms and $k-1$ for
the boundary ones.

The following property of the heat kernel coefficients
\cite{Gilkey:1994} will be useful in the calculations. Let us
define a shifted operator
\begin{equation}
L_\epsilon = L-\epsilon Q \,.\label{Lepsilon}
\end{equation}
Then
\begin{equation}
\left. \frac{d}{d\epsilon} \right\vert_{\epsilon =0} {\rm Tr} (
\exp (-t L_\epsilon )) =t{\rm Tr} (Q\exp (-tL)) \,.\label{var1}
\end{equation}
By expanding both sides of this equation in a power series of $t$
one obtains: 
\begin{equation}
\left. \frac{d}{d\epsilon} \right\vert_{\epsilon =0} a_{k+2}
(1,L_\epsilon )=a_k(Q,L) \,.\label{var2}
\end{equation}
All geometric quantities (metric, effective connection, boundary
conditions, etc.) corresponding to $L_\epsilon$ are the same as
for the unperturbed operator $L$ except for the potential $E$
which receives a shift,
\begin{equation}
E_\epsilon = E+\epsilon Q. \label{Eeps}
\end{equation}
Therefore, variation of the heat kernel coefficients w.r.t.
$\epsilon$ is equivalent to the variation of $E$. Equation
(\ref{var2}) together with the heat kernel coefficients for scalar
smearing function presented in Appendix \ref{hkapp} allow to
calculate the coefficients $a_k(Q,L)$ for $k=0,1,2,3$:
\begin{eqnarray}
&&a_0(Q,L)=(4\pi)^{-n/2}\iM \,{\rm tr}\,(Q).\label{Qa0bou} \\
&&a_1(Q,L)={\frac 14}(4\pi)^{-(n-1)/2}\idM
     \,{\rm tr}\, (\chi Q). \label{Qa1bou} \\
&&a_2(Q,L)=\frac 16 (4\pi)^{-n/2}\left\{ \iM
     \,{\rm tr}\,(6QE+QR) \right. \nonumber \\
&&\qquad\qquad \left. +\idM \,{\rm tr} \,
    (2Q L_{aa} +12QS +3 \chi Q_{ ;n})
     \right\} .\label{Qa2bou} \\
&&a_3(Q,L)=\frac 1{384}(4 \pi )^{
       -(n-1)/2}  \idM
       {\rm tr} \big\{ Q( -24 E + 24 \chi E \chi \nonumber\\
&&\qquad\qquad +48 \chi E + 48 E\chi
                     -12 \chi_{ :a} \chi_{:a} + 12 \chi_{:aa}
              -6 \chi_{:a}\chi_{:a}\chi +
              16\chi  R
\nonumber \\ &&\qquad\qquad + 8 \chi R_{anan} +192 S^2 + 96 L_{aa}
S + (3+10\chi )L_{aa}L_{bb} \nonumber \\ &&\qquad\qquad +(6-4\chi
) L_{ab}L_{ab} ) + Q_{;n}(96 S +192 S^2)
              +24 \chi Q_{ ;nn} \big\}.\label{Qa3bou}
\end{eqnarray}

Since the boundary terms in $a_6(1,L)$ are not known, we have to
adopt a different strategy to calculate $a_4(Q,L)$. The volume
part of $a_4(Q,L)$ is already known \cite{hknobou}, so that we have to
define the boundary contributions only. First we have to write
down all possible local boundary invariants of dimension $3$ with
arbitrary coefficients. Boundary invariants are traces over ``internal''
indices of local polynomials constructed from $R$, $E$, $\chi$, $\Omega$,
$L$, $S$ and $Q$ and from their derivatives. All $a,b,c,\dots$ indices
must be contracted in pairs. Note, that the normal index $n$ must not
be contracted. This reflects specific symmetry of the spectral problem
in the presence of a boundary which selects direction of the normal. 
Next, we have to use various properties of
the heat kernel expansion to define these constants. In the
particular case $Q=If$ we should restore the known result
(\ref{a4bou}). Of course, for a matrix valued $Q$ there are
considerably more different invariants than in the scalar case
since one has to take into account non-commutativity of $Q$ with
$E$, $\chi$, $\Omega$ etc. Therefore, for example, the invariants
$\tr (QE)$ and $\tr (Q\chi E \chi)$ are different although they
coincide in the limit $Q=If$. To restrict the number of invariants
(and the computational complexity) from now on we consider the
case
\begin{equation}
S=0,\qquad L_{ab}=0 \label{restrict}
\end{equation}
only.

Since the Riemann tensor does not have internal (spinorial or
gauge) indices, it commutes with $Q$ and $\chi$. Therefore, the
particular case $Q=If$ allows to restore all curvature dependent
boundary terms in $a_4$:
\begin{eqnarray}
&&a_4(Q,L)[{\mbox{boundary}}]= \frac 1{360} (4\pi)^{-n/2} \idM \tr
\left( Q(12R_{;n} +
30\chi R_{;n})\right.\nonumber \\
&&\qquad\qquad\qquad\qquad \left.+ 30 Q_{;n}\chi R  +
O(R^0)\right) \,. \label{R-terms}
\end{eqnarray}

To control the invariants containing $E$ and $\Omega$ we use the
property \cite{Branson:1990,Gilkey:1994} that the constants
appearing in front of all invariants depend on the dimension of
the manifold only through an overall factor $(4\pi )^{-n/2}$. This
property makes it possible to use low dimensional particular case
calculations to define the heat kernel coefficients in arbitrary
dimension.


\subsection{Particular case calculations}\label{partcase}

To define the terms in $a_4$ which depend on $\chi$, $E$ and
$\Omega$ it is enough to consider the case of the simplest
geometry $M=\mathbb{R}_+\times \mathbb{R}^{n-1}$ with flat metric.
It is important that the mass dimension of the boundary integrand
in $a_4(Q,L)$ is three. Therefore, terms containing both $E$ and
$\Omega$ cannot appear. For this reason, $E$ and $\Omega$ terms
can be considered separately. As well, $\Omega_{ab}$ cannot enter
the invariants since there is no rank two antisymmetric tensor of
dimension one. Consequently, we may restrict ourselves to the case
\begin{equation}
\omega_n=0 \,.\label{omegan}
\end{equation}
To simplify the calculations we also impose
\begin{equation}
\chi=const. \label{chiconst}
\end{equation}
Then $\chi_{:a}$ will be represented by a commutator
$[\omega_a,\chi]$, and $\Omega_{an}= -\partial_n\omega_a$. These
two invariants are independent on the boundary.

We shall need a bi-local heat kernel $K(x,z;t)$ which is defined
as a solution of the heat equation
 \begin{equation}
 (\partial_{t} + L)\, K(x,z;t) = 0 \label{fullkernel}
 \end{equation}
 with the initial condition
 \begin{equation}
 K(x,z;0) = \delta(x,z) \,.\label{initial}
 \end{equation}
Because of the restrictions (\ref{restrict}) and (\ref{omegan})
the boundary conditions simplify to
\begin{equation}
\Pi_- K(x,z;t)\BB =0\,,\quad \partial_n  \Pi_+ K(x,z;t)\BB =0 .\,,
\label{mixedkernel}
\end{equation}

This bi-local kernel is related to the ``localised'' one (cf.
(\ref{defhk})) by the equation:
\begin{equation}
K(Q,L,t)=\iM \tr \left( Q(x) K(x,x;t) \right) \,.\label{kernels}
\end{equation}
We stress, that $K(x,x;t)$ is a distribution.

The fundamental solution of ``free'' heat conduction equation
\begin{equation}
(\partial_{t} - \partial^2_x)\, K_0(x,z;t) = 0
 \label{freeheat}
\end{equation}
on $M=\mathbb{R}^n$ is well known:
\begin{equation}
K_0 (x^{a},x^{n}, y^{a}, y^{n}; t) = (4\pi t)^{-n/2} {\exp\Bigl(-
 \frac{\sum_{a=1}^{n-1}(x^{a}-y^{a})^2 +
(x^{n}-y^{n})^2}{4t}\Bigr)}  .
\end{equation}
From this kernel one can construct a solution of (\ref{freeheat})
on $M=\mathbb{R}_+\times \mathbb{R}^{n-1}$ which satisfies the
conditions (\ref{mixedkernel}) at $x^n=0$:
\begin{equation}
K_{\chi} (x, z; t) =
 K_0 (x^{a}, x^{n}, z^{a}, z^{n}; t) + \chi
K_0 (x^{a}, x^{n}, z^{a}, -z^{n}; t) .
\end{equation}

The operators which we consider in this section can be represented
in the following form:
\begin{equation}
L = - \partial^2 - P,
\end{equation}
where $P$ is a first or zeroth order differential operator.

It is easy to show that the kernel $K(x,z;t)$ defined by the
equation
\begin{equation}
K(x,z;t) = K_{\chi} (x,z;t) + \int_0^t d\tau \int_{M} dy\,
K_{\chi}(x,y; t-\tau) P(y) K(y,z; \tau) \label{inteq}
\end{equation}
satisfies both the full heat equation (\ref{fullkernel}) and the
boundary conditions (\ref{mixedkernel}).

The equation (\ref{inteq}) admits a solution\footnote{On
manifolds without boundary a similar expansion served as a starting
point for the covariant perturbation theory of ref.\ \cite{Barvinsky:uw}.
On manifolds with boundary similar perturbation series were used
\cite{Bordag:2001fj} to evaluate dependence of the heat kernel on the
boundary conditions.} in terms of power series in $P$:
\begin{eqnarray}
&&K(x,z;t) = K_{\chi}(x,z;t)  \nonumber \\
&&\quad + \sum_{p=1}^{\infty} \int_{0}^{t} d\tau_{p}
 \int_{0}^{\tau_{p}} d\tau_{p-1} \ldots
 \int_{0}^{\tau_{2}} d\tau_{1}  \int_M dy_p
 \ldots
 \int_M dy_1 \nonumber \\
&&\quad \times K_{\chi} (x,y_{p};t-\tau_{p}) P(y_{p})
 K_{\chi} (y_{p},y_{p-1};\tau_{p}-\tau_{p-1}) \ldots
 P(y_{1}) K_{\chi} (y_{1},z; \tau_{1})
\label{expansion}
 \end{eqnarray}
A remarkable feature of (\ref{expansion}) is that only a finite
number of terms contribute to each $a_k(Q,L)$ for fixed $k$. This
equation will be used to calculate the heat kernel expansion for
two particular choices of $L$ (see below).

We start with a one-dimensional example
\begin{equation}
L_1=-\partial_x^2 -E(x), \label{L1}
\end{equation}
so that $P_1(y) = E(y)$. $E$ is an arbitrary matrix valued
function on $M=\mathbb{R}_+$. Details of the calculation can be
found in Appendix \ref{partapp}. To the first order in $E$ we
have:
\begin{eqnarray}
&&K(Q,L_1,t)= \frac{t^{\frac{1}{2}}}{(4\pi)^{\frac{1}{2}}}
\int_0^\infty dx\, {\rm tr } \left( Q
E +\frac{t}{6} Q E_{;\mu}^{\:\:\mu} \right) \nonumber \\
&&\qquad +\frac{t}{(4\pi)^0}  {\rm tr} \, Q \Bigl(
-\frac{1}{16}(E-\chi E\chi) +
\frac{1}{8} (\chi E + E\chi)  \Bigr)_{x=0}  \nonumber\\
&&\qquad +\frac{t^{\frac{3}{2}}}{(4\pi)^{\frac{1}{2}}}  {\rm tr}
 \left[ Q
\Bigl(\frac{1}{12}(E_{;n} + \chi E_{;n} \chi) +\frac{1}{4}(\chi
E_{;n} + E_{;n}\chi) \Bigr) \right. \nonumber\\
&&\qquad \left. + Q_{;n} \Bigl( -\frac{1}{12}(E - \chi E \chi) +
\frac{1}{4} (\chi E + E \chi) \Bigr)\right]_{x=0}  + O(t^2)\,.
 \label{Eterm}
\end{eqnarray}
The first and the second lines of (\ref{Eterm}) can serve as a
consistency check of our calculations (cf. (\ref{Qa2bou}),
(\ref{Qa3bou})). The rest of (\ref{Eterm}) defines uniquely all
terms in the boundary part of $a_4(Q,L)$ containing $E$ subject to
the restrictions (\ref{restrict}). Indeed, no other invariants of
dimension 3 can appear. For example, $\tr (QE_{:a})$ is not
allowed since it contains a non-contracted tangential index.
$\chi_{;n}$ cannot appear since $\chi$ is defined on the boundary
only and, hence, may be differentiated only tangentially.

In order to define the term in $a_4(Q,L)$ containing $\Omega$
and/or tangential derivatives $\chi$ we consider a two dimensional
example $M=\mathbb{R}\times \mathbb{R}_+$ and $L_2=-(\partial_\mu
+\omega_\mu )^2$. In addition to (\ref{omegan}) we also suppose
$\partial_a\omega_a=0$. Clearly, this condition does not exclude
any relevant invariant. Then $P_2(y) = 2 \omega_a (y^n)\partial_a
+ \omega_a^2(y^n)$. This time we need the first and second order
terms in the expansion (\ref{expansion}). All necessary technical
tools can be found again in Appendix \ref{partapp}. The result
reads:
\begin{eqnarray}
&&K(Q,L_2,t)= \frac{t^{\frac{1}{2}}}{{(4\pi)}^{\frac{1}{2}}}
\int\limits_{\partial M} dx\,{\rm tr} \, Q\Bigl(-\frac{1}{32}
\chi_{:a}\chi_{:a} + \frac{1}{32} \chi_{:aa}
- \frac{1}{64} \chi_{:a}\chi_{:a}\chi \Bigr) \nonumber \\
&&\qquad +\frac{t}{4\pi} \int\limits_M d^2x\,{\rm tr} \, Q
\frac{1}{12}
\Omega_{\mu\nu}\Omega^{\mu\nu} \nonumber\\
&&\qquad +\frac{t}{4\pi} \int\limits_{\partial M}dx \, {\rm tr}
\left\{ Q\Bigl( \frac{1}{20}\chi\chi_{:a}\Omega_{an} +
\frac{1}{30}\chi_{:a}\Omega_{an}\chi +
 \frac{1}{20}\Omega_{an}\chi\chi_{:a} \right. \nonumber\\
&&\qquad -\frac{1}{30}\chi\Omega_{an}\chi_{:a} +\frac{1}{60}
[\chi\Omega_{an}\chi, \chi_{:a}] + \frac{3}{20} [\chi_{:a},
\Omega_{an}] + \frac{1}{12} [\chi , \Omega_{an:a}]
 \Bigr)   \nonumber\\
&&\qquad + \left. Q_{;n} \Bigl( -\frac{1}{20} \chi_{:a}\chi_{:a} +
 \frac{1}{12}\chi_{:aa} - \frac{1}{60}\chi_{:a}\chi_{:a}\chi
 \Bigr) \right\} +
 O(t^{\frac{3}{2}})\,. \label{omegaterm}
\end{eqnarray}

Next we collect individual contributions contained in
(\ref{R-terms}), (\ref{Eterm}) and (\ref{omegaterm}) to obtain:
\begin{eqnarray}
&&a_4(Q,L)=\frac 1{360} (4 \pi )^{
       -n/2} \Big\{ \iM \,{\rm tr}\,
       \big\{ Q(60{E_{ ; \mu}}^\mu+60 R E+180E^2 \nonumber \\
&&\qquad\qquad +30 \Omega_{\mu\nu}
      \Omega^{\mu\nu} +12 {R_{;\mu}}^\mu +
      5 R^2-2R_{\mu\nu}R^{\mu\nu}+2R_{\mu\nu\rho\sigma}R^{\mu\nu\rho\sigma})
\big\} \nonumber \\
&&\qquad\qquad+ \idM \,{\rm tr} \,
      \big\{ Q \{ 30 E_{;n} + 30 \chi E_{;n} \chi +
      90 \chi E_{;n} + 90 E_{;n}\chi  \nonumber\\
&&\qquad\qquad +18\chi\chi_{:a}\Omega_{an} + 12
\chi_{:a}\Omega_{an}\chi +
  18 \Omega_{an}\chi\chi_{:a}  -  12 \chi\Omega_{an}\chi_{:a}
  \nonumber\\ &&\qquad\qquad
         + 6 [\chi\Omega_{an}\chi, \chi_{:a}] + 54
[\chi_{:a}, \Omega_{an}] + 30 [\chi , \Omega_{an:a}] + 12 R_{ ;n}
+ 30 \chi R_{ ;n} \} +
 \nonumber\\ &&\qquad\qquad
       + Q_{ ;n}(- 30 E + 30 \chi E \chi + 90 \chi E
+ 90 E \chi - \nonumber \\
&&\qquad\qquad
 -18 \chi_{ :a} \chi_{ :a}   + 30 \chi_{:aa}
 - 6 \chi_{:a}\chi_{:a}\chi + 30
\chi R )+  30 \chi {Q_{ ;\mu}}^{\mu n}  \big\} \Big\} .
\label{Qa4bou}
\end{eqnarray}
Let us remind that we have imposed a restriction (\ref{restrict})
on boundary conditions and on the extrinsic curvature of the
boundary.


\section{Calculation of the anomaly}


\subsection{Dimension two}
The remaining part of the calculation is rather simple. One has to
substitute (\ref{Qa2bou}) with (\ref{B9}) - (\ref{Om}), (\ref{B11})
and (\ref{bagbc}) in (\ref{chian}). The result reads:
\begin{equation}
\mathcal{A}=-\frac 1\pi \int_M d^2x \sqrt{g}\, \tr \left( \varphi
\Bigl(\frac{1}{2} \epsilon^{\mu\nu} (V_{\mu\nu} + [A_\mu,A_\nu ])
- D_\mu A^{\mu} \Bigr) \right) \,.\label{Ain2}
\end{equation}
Surprisingly, there is no ``genuine'' boundary contribution here,
and this is our main result in two dimensions. For constant $\varphi$
it is consistent with an earlier calculation \cite{Wipf:1994dy}.
The volume term is very well known (see, e.g. \cite{books}).
Note, that in two dimensions axial vectors can be transformed to
vectors, and, therefore, the whole anomaly may be generated by just the
first term under the integral in (\ref{Ain2}).


\subsection{Dimension four}
Chiral anomaly in four dimensions is calculated the same way
except for that we have to use eq.\ (\ref{Qa4bou}). Now the
anomaly contains two contributions. 
\begin{equation}
\mathcal{A}=\mathcal{A}_v+\mathcal{A}_b \,. \label{Ain4}
\end{equation}
The volume part
\begin{eqnarray}
    \label{B16}
&&\mathcal{A}_v=
    \frac{-1}{180 \, (2 \pi)^{2} } \int_M \, d^{4}x \sqrt{g} \,
    {\rm tr} \varphi
     \Bigl( - 120 \, [D_{\mu}V^{\mu\nu},A_\nu] \nonumber\\
&&\qquad\qquad  +60 \, [D_{\mu}A_\nu,V^{\mu\nu}]  
   - 60 \, D_{\mu}D^{\mu}D_{\nu}A^\nu
   + 120 \, \{\{D_{\mu}A_\nu , A^\nu\} , A^\mu\}\nonumber\\
&&\qquad\qquad   + 60 \, \{D_{\mu} A^\mu,A_\nu A^\nu \}  
    + 120 \, A_\mu D_{\nu}A^\nu A^\mu
   + 30 \, [[A_\mu , A_\nu],A^{\mu\nu}]   \nonumber \\
&&\qquad\qquad    + \epsilon_{\mu\nu\rho\sigma}\
   \{ - 45 \, i\, V^{\mu\nu}V^{\rho\sigma}  
    + 15 \, i \, A^{\mu\nu}A^{\rho\sigma}
   - 30 \, i \, (V^{\mu\nu}A^{\rho} A^{\sigma} 
   + A^{\mu} A^{\nu} V^{\rho\sigma})
   \nonumber  \\
&&\qquad\qquad   - 120 \, i \, A^{\mu} V^{\nu\rho} A^{\sigma}
     + 60 \, i \, A^{\mu} A^{\nu} A^{\rho} A^{\sigma}\}  
   - 60 \, (D_{\sigma}A_\nu) R^{\nu\sigma} 
  + 30 \, (D_{\mu}A^\mu) R \nonumber \\
&&\qquad\qquad   
   - \frac{15i}{8} \epsilon_{\mu\nu\rho\sigma} 
\, R^{\mu\nu}{}_{\eta \theta}
    R^{\rho\sigma \eta \theta} 
    \Bigr )  
\end{eqnarray}
is known\footnote{The flat space part of (\ref{B16}) can be found e.g. in
\cite{Andrianov:1983fg}. The term with $R\star R$ follows from the local
index theorem and has a very long history \cite{Eguchi:jx}. We are not aware
of any works which considered chiral anomaly for non-abelian axial vector
field in curved space, so that the terms $(DA)R$ have a chance to be new.
There are certain similarities between (\ref{B16}) and chiral anomaly
in the Riemann-Cartan space \cite{RCchi}.}.

The boundary part
\begin{eqnarray}
&&\mathcal{A}_b=
    \frac{-1}{180 \, (2\pi)^2} \int_{\partial M} \, d^{3}x \sqrt{h}
    \,   {\rm tr} \Bigl(
    12 \, i \, \epsilon^{abc} \, \{A_b, \varphi \} D_a A_c
\nonumber \\
 &&\qquad\qquad   + 24 \{\varphi, A^a \} \{A_a, A_n\}  
    -60 \, [A^a, \varphi] (V_{na} - [A_n, A_a]) \nonumber\\
&&\qquad\qquad  + 60 (D_n \varphi ) D_{\mu} A^{\mu}  \Bigr) \label{banomaly}
   \end{eqnarray}
is new. It has been derived under two restrictions (\ref{restrict}).
Note, that in the present context, the first condition ($S=0$) actually
follows from the second one ($L_{ab}=0$) due to (\ref{B11}).
We shall analyse physical consequences of (\ref{banomaly}) in a future
publication.


\section{Conclusions}
We have calculated boundary contributions to chiral anomaly for
local (bag) boundary conditions in two and four dimensions
(in four dimensions we have supposed that the boundary is totally
geodesic, $L_{ab}=0$). As a by-product we obtained explicit expressions
for several heat kernel coefficients with mixed boundary conditions
and with a matrix-valued smearing function. These heat kernel expressions
are rather universal. By choosing a bit different expressions for
$Q$ and for the fields (\ref{B9}) - (\ref{Om}) one can easily
extend our results to other anomaly-like expression relevant for
hadron physics (see, e.g., \cite{otheran}) and, probably,
even to supersymmetry \cite{Lindstrom:fr}.

We have found no specific boundary contributions to the anomaly
in two dimensions. In four dimensions there are boundary
terms in the anomaly, which must have important physical
consequences both in hadron physics and in the standard model.

Our present paper is the first one which treats local chiral anomaly
in the presence of boundaries
in generic background vector and axial vector fields. Therefore,
our results may be improved or extended in many directions.

It is clear that in order to bring our results closer to physical
applications we have to lift the restriction $L_{ab}=0$ in four
dimensions (which excludes, for example, spherical boundaries in flat 
space). This is just a technical problem which can be solved by
the same methods as presented above. Another problem is to extend
our results beyond the linear order in the chiral transformation
parameter. Such an extension requires chirally transformed boundary
conditions (\ref{chibag}). We are going to address these two
problems in the near future.

Brane-world and domain wall configurations lead to interactions
confined at a singular surface.
Mathematically such interactions are described at the one-loop
level by some ``gluing conditions'' which relate
boundary values of the functions and their
normal derivatives on two sides of the singular surface. Such cases
may be treated by the same methods as presented in this paper.
Moreover, a lot of important information on the heat kernel
expansion for gluing conditions is contained in the heat
kernel expansion for the boundary conditions case
(see, e.g., \cite{doubling}). 

\section*{Acknowledgements}

We are grateful to Michael Bordag, Peter
Gilkey, Yu.~V.~Novozhilov and Peter van Nieuwenhuizen for useful comments and
stimulating discussions. This work has been supported in part by
the DFG project BO 1112/12-1 (D.V.V.) and by the
Ostpartnerschaften program of Leipzig University (V.N.M.). V.N.M.
thanks Michael Bordag for his hospitality
in Leipzig.

\appendix

\section{Heat kernel expansion with mixed boundary conditions}\label{hkapp}

Here we give explicit expressions for the heat kernel coefficients
$a_k$, $k=0,1,2,3,4$ for an operator of Laplace type subject to
mixed boundary conditions \cite{Branson:1990} (see also
\cite{Vassilevich:we} for minor corrections in $a_4$), $f$ is a
scalar function.

\begin{eqnarray}
&&a_0(f,L)=(4\pi)^{-n/2}\iM \, {\rm tr}\, (f) \label{a0bou} \\
&&a_1(f,L)={\frac 14}(4\pi)^{-(n-1)/2}\idM
     \, {\rm tr} \, (\chi f). \label{a1bou} \\
&&a_2(f,L)=\frac 16 (4\pi)^{-n/2}\left\{ \iM
    \, {\rm tr} \, (6fE+fR) \right. \nonumber \\
&&\qquad\qquad \left. +\idM \, {\rm tr} \, (2fL_{ aa}+3 \chi f_{
;n}+12fS)
     \right\} .\label{a2bou} \\
&&a_3(f,L)=\frac 1{384}(4 \pi )^{
       -(n-1)/2}  \idM \, {\rm tr} \, \big\{ f(96 \chi E+16\chi  R \nonumber\\
&&\qquad\qquad+8f \chi R_{anan}
       +(13 \Pi_{+}-7 \Pi_{ -})L_{ aa}L_{ bb}+(2 \Pi_{ +}+10
      \Pi_{-})L_{ab}L_{ ab}\nonumber\\
&&\qquad\qquad+96SL_{ aa}
      +192S^2
       -12 \chi_{ :a} \chi_{:a})+f_{ ;n}
      ((6 \Pi_{ +}+30 \Pi_{ -})L_{
        aa}\nonumber\\
&&\qquad\qquad+96S)+24 \chi f_{ ;nn} \}.\label{a3bou}\\
&&a_4(f,L)=\frac 1{360} (4 \pi )^{
       -n/2} \big\{ \iM \, {\rm tr}\,
       \{ f(60{E_{ ; \mu}}^\mu+60 R E+180E^2 \nonumber \\
&&\qquad\qquad +30 \Omega_{\mu\nu}
      \Omega^{\mu\nu} +12 {R_{;\mu}}^\mu +
      5 R^2-2R_{\mu\nu}R^{\mu\nu}+2R_{\mu\nu\rho\sigma}R^{\mu\nu\rho\sigma})
\} \nonumber \\
&&\qquad\qquad+ \idM \, {\rm tr}\,
      \big\{ f \{ (240 \Pi_{ +}-120 \Pi_{ -})E_{ ;n}\nonumber\\
&&\qquad\qquad+(42 \Pi_{
     +}-18 \Pi_{ -}) R_{ ;n}
     +24L_{ aa:bb}+120EL_{
       aa}\nonumber\\
&&\qquad\qquad +20 R L_{ aa}+4R_{ an an}L_{bb}
      -12R_{ an    bn}L_{ ab}+4R_{
       ab    cb}L_{ ac}\nonumber\\
&&\qquad\qquad +\frac 1{21} \{ (280 \Pi_{ +}+40 \Pi_{ -})L_{
aa}L_{ bb}L_{ cc}
    +(168 \Pi_{ +}\nonumber\\
&&\qquad\qquad -264 \Pi_{ -})L_{ ab}L_{ ab}L_{ cc}+(224 \Pi_{
+}+320 \Pi_{ -})L_{ ab}L_{ bc}L_{ac} \} \nonumber \\
&&\qquad\qquad +720SE+120S R +144SL_{
aa}L_{ bb}+48SL_{ ab}L_{ ab} \nonumber \\
 &&\qquad\qquad+480S^2L_{ aa}+480S^{
3}+120S_{ :aa}+60 \chi  \chi_{ :a} \Omega_{ an}-12 \chi_{ :a}
\chi_{
:a}L_{ bb} \nonumber \\
&&\qquad\qquad -24 \chi_{ :a} \chi_{ :b}L_{ ab}-120 \chi_{ :a}
\chi_{ :a}S \} +f_{ ;n}(180 \chi E+30 \chi  R
 \nonumber \\
&&\qquad\qquad +\frac 17 \{ (84 \Pi_{ +}-180 \Pi_{ -})L_{ aa}L_{
bb}+(84 \Pi_{
+}+60 \Pi_{ -})L_{ ab}L_{ ab} \} \nonumber \\
&&\qquad\qquad +72SL_{ aa}+240S^2-18 \chi_{ :a} \chi_{ :a})+f_{
;nn}(24L_{
aa}+120S) \nonumber \\
&&\qquad\qquad \left. +30 \chi {f_{ ;\mu}}^{\mu n}  \big\}
\right\} . \label{a4bou}
\end{eqnarray}

The coefficient $a_5$ can be found in \cite{Branson:1999jz}. Here
we present only those terms which enter (\ref{var2}) with $k=3$.
Namely, we put $f=1$ and neglect all terms which do not contain
$E$.
\begin{eqnarray}
&&a_5(1,L)=\frac{1}{5760}(4\pi)^{-(m-1)/2} \idM {\rm tr} \left(
360 \chi E_{;nn} + 1440 E_{;n}S
\right.\nonumber \\
&&\qquad\qquad +720 \chi E^2 + 180 \chi_{:aa} E + 240 \chi R E
-120 \chi R_{nn} E + 2880 ES^2 \nonumber\\
&&\qquad\qquad +(270 - 180\chi ) L_{aa} E_{;n} + 1440 L_{aa}SE
+(45 +150\chi )L_{aa}L_{bb} E \nonumber \\
&&\qquad\qquad + (90 -60\chi )L_{ab}L_{ab} E -180 (E^2 -\chi E
\chi E) - 180 \chi_{:a}\chi_{:a} E
\nonumber \\
&&\qquad\qquad \left. -90 \chi \chi_{:a}\chi_{:a} E \right) +
O(E^0) .\label{a5bou}
\end{eqnarray}


\section{Particular case calculation: details}\label{partapp}
Let us consider in detail calculations of the expansion
(\ref{Eterm}) for the operator (\ref{L1}). First we use the
(\ref{expansion}) to represent the first order terms in $E$ of
$K(Q,L_1;t)$ as
\begin{eqnarray}
&&\frac 1{4\pi} \int\limits_0^t d\tau \int\limits_0^\infty dx
\int\limits_0^\infty dy \frac 1{\sqrt{\tau}\sqrt{t-\tau}} \tr
\left\{ Q(x)\left( e^{ -\frac{(x-y)^2}{4\tau}} +\chi e^{
-\frac{(x+y)^2}{4\tau}}
\right)  \right. \nonumber \\
&&\qquad \times \left. E(y)\left( e^{ -\frac{(x-y)^2}{4(t-\tau)}}
+\chi e^{ -\frac{(x+y)^2}{4(t-\tau)}} \right)
 \right\} \label{Elin}
\end{eqnarray}
Next we integrate over $\tau$ with the help of the relation
\begin{equation}
\int_0^t d \tau \frac{\exp(-\frac{a^2}{\tau})
\exp(-\frac{b^2}{t-\tau}) }{\sqrt{\tau}\sqrt{t-\tau}}= \pi \cdot
\mathrm{erfc}\Biggl(\frac{|a|+|b|}{\sqrt{t}}\Biggr) .
\label{tauint1}
\end{equation}
The parameters $a$ and $b$ are equal either to $(x-y)/2$ or to
$(x+y)/2$. Note, that ``reflected'' terms depending on $(x+y)$ are
always multiplied by $\chi$.

It is convenient to consider $3$ different types of contributions separately.
\begin{enumerate}
\item Terms without $\chi$. \\
In this case $a=b=(x-y)/2$. Let us change the variables
\begin{eqnarray}
&&x=k+r\sqrt{t} ,\qquad y=k \quad {\mbox{for}}\ x>y \nonumber\\
&&y=k+r\sqrt{t} ,\qquad x=k \quad {\mbox{for}}\  x<y .
\end{eqnarray}
In both cases $k, r \in [0, +\infty[$. Then we use a small $t$
expansion:
\begin{equation}
\int_{0}^{+\infty} dr \,{\rm erfc}(r) f(r\sqrt{t}) =
\frac{f(0)}{\sqrt{\pi}} + \frac{f^{\,'}(0)\sqrt{t}}{4} +
\frac{f^{''}(0) t}{6\sqrt{\pi}} + \dots
\end{equation}
which is valid for any smooth function which decays sufficiently
fast at infinity. We obtain:
\begin{eqnarray}
&&\int_{0}^{+\infty} dx \int_{0}^{+\infty} dy \, {\rm erfc}
\biggl(\frac{|x-y|}{\sqrt{t}}\biggr) {\rm tr}\frac{Q(x)E(y)}{4}
\nonumber \\
&&=\int_{0}^{+\infty} dk \int_{0}^{+\infty} dr \, {\rm erfc} (r)
{\rm tr}\frac{\bigl(Q(k+\sqrt{t}r) E(k) + Q(k)E(k+\sqrt{t}r)
\bigr)}{4} \nonumber \\
&&=\frac{t^{\frac{1}{2}}}{2\sqrt{\pi}}\int_{0}^{+\infty} dy\, {\rm
tr} \,\left( Q(y)E(y)   + \frac{t}{6} Q(y) E\, ''(y)  \right)
- \frac{t}{16} {\rm tr} \,  Q(0) E(0) \nonumber\\
&&\quad +\frac{t^{\frac{3}{2}}}{2\sqrt{\pi}} {\rm tr}
\biggl(\frac{1}{12} Q(0) E\,'(0)  - \frac{1}{12} Q\,'(0)E(0)
\biggr) + O(t^2)\,. \label{term1}
\end{eqnarray}

\item Terms with two $\chi$. In this case $a=b=(x+y)/2$.
We change the variables:
\begin{equation}
x=r\sqrt{t}\cos\phi \,,\qquad
 y=r\sqrt{t}\sin\phi \,,
\end{equation}
so that $r\in [0,\infty [$, $\phi \in [0,\pi/2 ]$. We use
(\ref{tauint1}) and then integrate over $r$ and $\phi$ assuming $t$ is
small . The resulting asymptotic expansion reads
\begin{eqnarray}
&&\int_{0}^{+\infty} dx \int_{0}^{+\infty} dy \, {\rm erfc}
\biggl(\frac{|x+y|}{\sqrt{t}}\biggr) {\rm tr} \,\frac{Q(x)\chi
E(y)\chi}{4}
=\frac{t}{16} {\rm tr} \, Q(0)\chi E(0)\chi \nonumber \\
&&\quad +\frac{t^{\frac{3}{2}}}{2\sqrt{\pi}} {\rm
tr}\Bigl(\frac{1}{12} Q\,'(0)\chi E(0)\chi + \frac{1}{12} Q(0)\chi
E\,'(0)\chi \Bigr)+ O(t^2) \label{term2}
\end{eqnarray}

\item Terms with single $\chi$. We have $a=(x-y)/2$, $ b=(x+y)/2$.
We use the variables:
\begin{eqnarray}
x=r\sqrt{t}(\cos\phi + \sin\phi) ,\qquad y=r\sqrt{t}\cos\phi \quad
{\mbox{for}} \ x>y;
\nonumber \\
y=r\sqrt{t}(\cos\phi + \sin\phi) ,  \qquad x=r\sqrt{t}\cos\phi
\quad {\mbox{for}} \ x<y.
\end{eqnarray}
Acting as above we obtain:
\begin{multline}
\int_{0}^{+\infty} dx \int_{0}^{+\infty} dy \, {\rm erfc}
\biggl(\frac{|x+y|+|x-y|}{2\sqrt{t}}\biggr)
{\rm tr} \,\frac{Q(x)\bigl(\chi E(y) + E(y)\chi\bigr)}{4} = \\
\frac{t}{8} \, {\rm tr}\, Q(0)\bigl(\chi E(0)+ E(0)\chi\bigr) +\\
\frac{t^{\frac{3}{2}}}{2\sqrt{\pi}}{\rm
tr}\Bigl(\frac{1}{4}Q(0)\bigl(\chi E\,'(0) +E\,'(0)\chi\bigr) +
\frac{1}{4} Q\,'(0)\bigl(\chi E(0) + E(0)\chi \bigr)
 \Bigr) + O(t^2) \label{term3}
\end{multline}
\end{enumerate}

The sum of (\ref{term1}), (\ref{term2}) and (\ref{term3}) yields
(\ref{Eterm}).

Calculations of the $\omega$ terms can be carried out in a similar
way. The following integral is useful:
\begin{multline}
\int_{0}^{t} dp \int_{0}^{p} d\tau \frac{\exp(-\frac{a^2}{t-p})
\exp(-\frac{b^2}{p-\tau}) \exp(-\frac{c^2}{\tau}) }{
\sqrt{t-p}\sqrt{p-\tau}\sqrt{\tau} } = \\= - 2 \pi^{3/2}\cdot
(|a|+|b|+|c|) \cdot \mathrm{erfc}
\Biggl(\frac{|a|+|b|+|c|}{\sqrt{t}}\Biggr) + 2 \pi \sqrt{t} \cdot
\exp\Biggl(-\frac{(|a|+|b|+|c|)^2}{t}\Biggr)
\end{multline}

\end{document}